\documentclass[aip, reprint, nofootinbib, nobibnotes]{revtex4-1}

\usepackage{amssymb,amsfonts,amsmath, mathtools}
\usepackage{graphicx}

\begin{document}

\title{Stepping and crowding of molecular motors: statistical kinetics from an exclusion process perspective}
\date{26 May 2014}

\author{Luca Ciandrini}
\email[To whom correspondence should be addressed.\\E-mail:]{luca.ciandrini@univ-montp2.fr}
\affiliation{Laboratoire de Dynamique des Interactions Membranaires Normales et Pathologiques UMR 5235, Universit\'e Montpellier II and CNRS, 34095 Montpellier Cedex 5, France}
\affiliation{Laboratoire Charles Coulomb UMR 5221, Universit\'e Montpellier II and CNRS, 34095 Montpellier, France}

\author{M. Carmen Romano}
\affiliation{Institute for Complex Systems and Mathematical Biology, SUPA, University of Aberdeen, King's College, AB24 3UE}
\affiliation{Institute of Medical Sciences, Foresterhill, University of Aberdeen, Aberdeen AB25 2ZD, United Kingdom}

\author{Andrea Parmeggiani}
\affiliation{Laboratoire de Dynamique des Interactions Membranaires Normales et Pathologiques UMR 5235, Universit\'e Montpellier II and CNRS, 34095 Montpellier Cedex 5, France}
\affiliation{Laboratoire Charles Coulomb UMR 5221, Universit\'e Montpellier II and CNRS, 34095 Montpellier, France}


\begin{abstract}
Motor enzymes are remarkable molecular machines that use the energy derived from the hydrolysis of a nucleoside triphosphate to generate mechanical movement, achieved through different steps that constitute their kinetic cycle. These macromolecules, nowadays investigated with advanced experimental techniques to unveil their molecular mechanisms and the properties of their kinetic cycles, are implicated in many biological processes, ranging from biopolymerisation (e.g. RNA polymerases and ribosomes) to intracellular transport (motor proteins such as kinesins or dyneins). 
Although the kinetics of individual motors is well studied on both theoretical and experimental grounds, the repercussions of their stepping cycle on the collective dynamics still remains unclear. Advances in this direction will improve our comprehension of transport process in the natural intracellular medium, where processive motor enzymes might operate in crowded conditions.  
In this work, we therefore extend the current statistical kinetic analysis to study collective transport phenomena of motors in terms of lattice gas models belonging to the exclusion process class. 
Via numerical simulations, we show how to interpret and use the randomness calculated from single particle trajectories in crowded conditions. Importantly, we also show that time fluctuations and non-Poissonian behavior are intrinsically related to spatial correlations and the emergence of large, but finite, clusters of co-moving motors.
The properties unveiled by our analysis have important biological implications on the collective transport characteristics of processive motor enzymes in crowded conditions.
\end{abstract}


\maketitle

\section*{INTRODUCTION}
The kinetic cycle of an enzyme, defined as the set of reactions and conformational changes necessary to accomplish its enzymatic activity, has an intrinsically stochastic nature. The aleatory character has significant repercussions on the kinetics of linear processive enzymes, such as motor proteins, and on the features of their stepping dynamics over a specific track.
The recent breakthroughs in high-resolution single-molecule techniques make it possible to observe and reliably quantify fluctuations in the completion time of individual enzymatic (stepping) cycles of isolated motors.  Statistical analysis of these variations allows then extracting information on key molecular features~\cite{schnitzer_statistical_1995, santos_renewal_2005, moffitt_methods_2010, moffitt_mechanistic_2010, chowdhury_stochastic_2013}. 

In the natural intracellular environment, however, many enzymatic processes occur in crowded conditions. Motor proteins moving along microtubules~\cite{leduc_molecular_2012}, multiple ribosomes translating an mRNA molecule~\cite{wolin_ribosome_1988}, or polymerases transcribing a DNA strand~\cite{klumpp_stochasticity_2008} are prominent examples in which the single motor stepping cycle driven by nucleotide hydrolysis (e.g. $ATP \rightarrow ADP +P_i$) can strongly depend on the presence of other motors moving on the same track. Motor stepping, indeed, depends also on the availability of free sites of the same track (cytoskeletal filaments and nucleic acids). These sites are necessary to catalyze nucleotide hydrolysis, which determines the motor mechanochemical cycle in interaction with its specific pathway~\cite{howard_mechanics_2001}.
Collective and cooperative phenomena of processive motors, on the other hand, have been largely studied in the theoretical  literature~\cite{chowdhury_physics_2005, chou_non-equilibrium_2011, bressloff_stochastic_2013} and in particular, in the field of driven lattice gases.
Most of these works describe motors as kinetically unstructured elements, despite the many intermediary mechano-chemical conformational changes underlying one single motor enzyme step. Some of them (see e.g.~\cite{aghababaie_universal_1999, nishinari_intracellular_2005, klumpp_effects_2008, garai_fluctuations_2009, ciandrini_role_2010}) explore also the role of the internal kinetic cycle of the motors on emerging collective properties.

A minimal representation of the stepping cycle of a motor involving the consumption of an energetic substrate $S$ (e.g., ATP or GTP) can be given by Michaelis-Menten enzymatic kinetics where  the catalysed chemical reaction and the enzyme spatial translocation are ``tightly-coupled'':
\begin{equation}
	\left( E+S \right)_x \xrightleftharpoons[{k_-}]{k_+ [S]} \left( ES \right)_x \xrightarrow[] {\gamma} \left(E+P\right)_{x+\ell}.
	\label{eq::michaelis}
\end{equation}
The substrate $S$ binds the enzyme $E$ with rate $k_+ [S]$ at position $x$ to first form a bound ``active  state'' $ES$ (where a backward rate $k_-$ is allowed) and then release the product $P$ with a rate $\gamma$. In this kinetic scheme, the completion of a stochastic stepping cycle corresponds to the advancement of the motor by a step of length $\ell$.  
The {\it dwell} or {\it residence} time $\tau$, i.e. the time needed between two outputs $P$, is a stochastic variable with fluctuations determined by the mechano-chemical cycle properties of the motor enzyme.

An important experimental quantity that is nowadays possible to reliably measure is the randomness parameter $r$~\cite{schnitzer_statistical_1995, moffitt_methods_2010,moffitt_mechanistic_2010}. This parameter quantifies the fluctuations occuring during the stepping cycle and is defined as the ratio of the variance of the dwell time $\tau$ to its squared average:
\begin{equation}
	r =\frac{\langle \tau ^2\rangle-\langle \tau \rangle^2}{\langle \tau \rangle^2} \;,
	\label{eq::randomness}
\end{equation}
where $\langle \cdot \rangle$ represents an average over a large number of events and experiments.  
Importantly, it has been experimentally shown that the randomness of many biological systems depends on the substrate concentration~\cite{schnitzer_kinesin_1997, visscher_single_1999}.
Moreover, it has been proved that in some conditions (see \cite{moffitt_methods_2010,moffitt_mechanistic_2010}) one can extract valuable information on the kinetic mechanisms of a single enzyme from the inverse of the randomness $n\equiv 1/r$, and associate it to the number of limiting steps of the kinetic cycle~\cite{schnitzer_statistical_1995,moffitt_methods_2010,moffitt_mechanistic_2010, chowdhury_stochastic_2013}. 

In this work, we present a model of interacting molecular motors endowed with internal kinetic cycles.
Our model brings insight into emergent features of motor enzyme collective dynamics from the perspective of statistical kinetics. We show that the randomness parameter $r$ is useful to study collective systems, although a different interpretation from single molecule studies is necessary. 
Crucially, the presence of (at least) an internal step in the single motor enzymatic kinetics can radically change the collective translocation of motors with excluded volume interactions with respect to models with interacting motors lacking an internal kinetic cycle. We thus provide a relation between the stochastic kinetics of a single tracer and the formation of extended clusters of motors moving along the same track. 
Interestingly, these properties can be related to the concentration of a biological fuel substrate (like ATP). The availability of the energy substrate, driving the enzymatic cycle and  translocation step, generates anomalous dynamics both at single molecule and at collective level. 
We then predict a universal mechanism controlled by the energy substrate availability that could regulate the formation of clusters of motors in {\it in vitro} and {\it in vivo} systems.  

\subsection*{A short overview on driven collective transport of Poissonian particles}
We first survey traffic models composed of molecular motors devoid of internal kinetic cycles (Poissonian walkers). This part is indeed important to contrast the novel behaviour induced by the presence of an internal state in the motor kinetics. Originally inspired by bio-polymerisation processes~\cite{macdonald_kinetics_1968}, the Totally Asymmetric Simple Exclusion Process (TASEP) describes the transport properties of particles, with (hard-sphere) exclusion interaction, moving in a preferred direction along a discrete lattice.
TASEP is the archetypal model used to study (biological and non-biological) unidimensional transport~\cite{chowdhury_physics_2005, chowdhury_statistical_2000, chou_non-equilibrium_2011}, theoretical aspects of probability theory~\cite{liggett_stochastic_1999,touchette_large_2009} and non-equilibrium statistical mechanics~\cite{chou_non-equilibrium_2011}. 
Recently, this standard model has posed the grounds for the development of theoretical tools describing the cooperative phenomena of processive molecular motors~\cite{lipowsky_random_2001, parmeggiani_phase_2003, parmeggiani_totally_2004}, with a good qualitative agreement when compared to state-of-art experiments~\cite{leduc_molecular_2012}.

\begin{figure}[h]
\begin{center}
\includegraphics[width=0.45\textwidth] {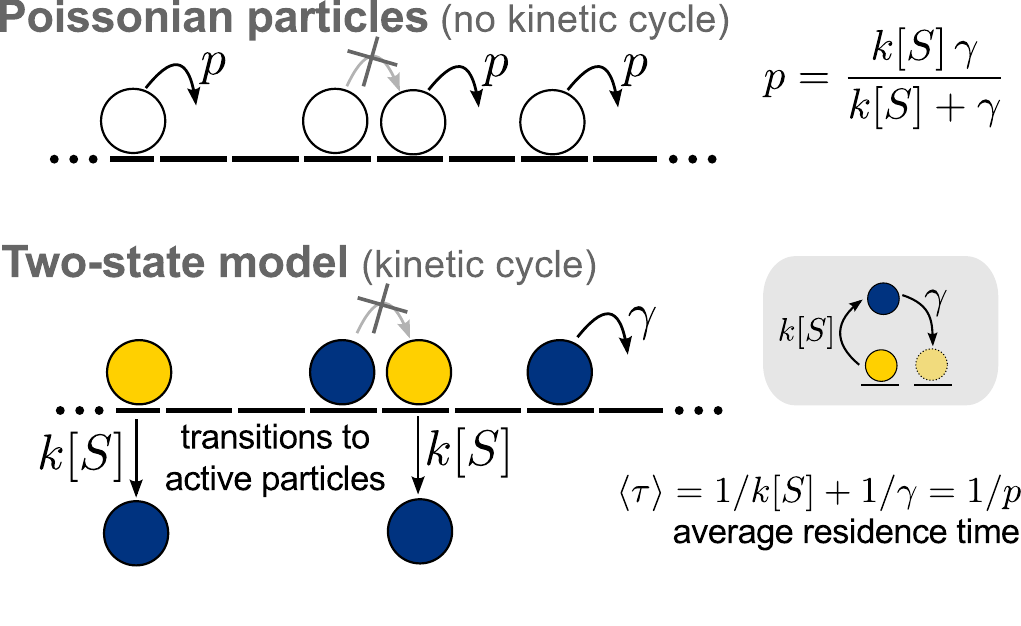}
\caption{Schematic representation of the collective models analysed. On top, unstructured particles moving along a filament following a Poissonian step with rate $p$, which depends on the substrate concentration through a Michaelis-Menten relation as described in the text.  Below, the two-state model presented in the main text, whose stepping cycle kinetics is represented in the dashed region. The average stepping time of an isolated particle is the same in the two models, allowing us to compare the two processes. \label{fig::model}}
\end{center}
\end{figure}

\begin{figure}[h!b]
\begin{center}
\includegraphics[width=0.5\textwidth]{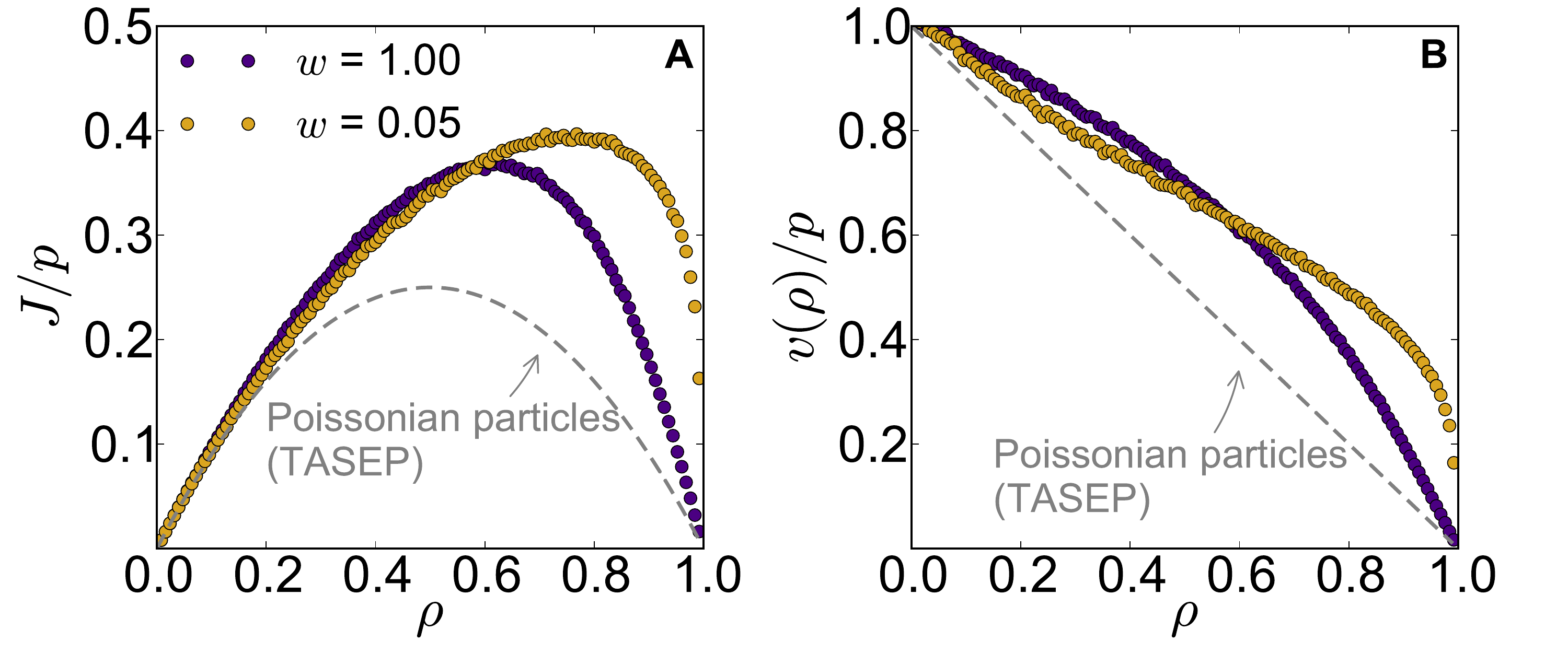}
\caption{Density-current relation ({\bf A}) and velocity ({\bf B}) of a lattice gas with and without particles' stepping cycle. The dashed gray line in {\bf A} represents $J/p=\rho(1-\rho)$ for a standard exclusion process, while the filled circles display the current of the two-state model at different substrate abundances $w$. Analogously, the dashed lines in panel {\bf B} show the velocity $v(\rho)/p=1-\rho$ typical of a set of Poissonian particles, and the filled circles are the outcome of the two-state model. Two different values of the substrate abundance are shown, $w=1$ (indigo) and $w=0.05$ (yellow).   \label{fig::comparison_current}}
\end{center}
\end{figure}

More specifically, TASEP consists of a unidimensional discrete lattice in which particles hop from one site to the next with probability per unit time $p$, provided that the entering site is empty (i.e. no more than one particle can occupy a site at the same time, see Fig.~\ref{fig::model}). The discrete lattice represents here the track (like a microtubule filament) along which the processive enzymes advance.
For particles lacking internal kinetic steps, the unique hopping rate $p$ depends on the substrate concentration providing the energy necessary for the movement of particles. Assuming a Michaelis-Menten-like dependence, it writes $p=k[S]\,\gamma/(k[S]+\gamma)$, where $\gamma$ represents the maximal hopping rate at saturating substrate concentration and $\gamma/k$ is the substrate concentration at which $p=\gamma/2$. 
When several particles occupy the lattice with a density $\rho$, the steady-state current $J$ (the number of particles passing through a site per unit time) is very well approximated (and exact in the limit of large systems) by the relation $J=\rho v(\rho)=p\,\rho(1-\rho)$, where $\rho$ is the density of particles on the lattice and $v(\rho)$ their velocity (see, e.g., \cite{derrida_exactly_1998}). This relationship is bounded in saturating substrate concentrations by $J=\gamma\,\rho(1-\rho)$, where $\gamma$ represents the maximal obtainable velocity at saturating substrate conditions. 
The current-density parabola typical of this driven particle model is shown in Figure~\ref{fig::comparison_current}A and the linear dependence of the velocity on the density is presented in Figure~\ref{fig::comparison_current}B (dashed lines). 

This simple model however presents severe limitations when compared to real motor proteins transport. On one hand, for tight coupled mechanisms at single molecule level, it is straightforward to show (starting from Eq.~[\ref{eq::randomness}]) that a Poissonian enzyme has always a randomness $r$ strictly equal to 1. A Poissonian walker is indeed characterised by exponential waiting time distributions with a typical time constant equal to $1/p$. Hence, without internal states, the randomness parameter $r$ introduced above in Eq.~[\ref{eq::randomness}] depends neither on the particles' hopping rate nor on the substrate concentration. This is in stark contrast to 
experimental results, for instance in the context of processive motor proteins\footnote{In these cases, the dwell time distribution can be written via a convolution of exponential distributions with typical times related to the microscopic limiting rate composing the kinetic cycle, and the randomness parameter $r$ becomes dependent on the substrate concentration.}~\cite{schnitzer_kinesin_1997,visscher_single_1999}.
Besides, at the collective level and in absence of internal states, changes in the substrate concentration only affect the timescale of the system leading to an invariant rescaled current  $J/p$, which remains the same for all values of the substrate concentration (dashed line in Fig.~\ref{fig::comparison_current}). This is not the case in presence of internal states as we will show below.
Taken together, Poissonian particles without internal kinetic steps hide interesting novel features of molecular motor collective transport behaviour.

\section*{RESULTS}
We describe how the particles' internal enzymatic kinetics affects the overall transport process at different levels, from mean quantities to temporal fluctuations and spatial dynamical effects. 
In a first stage, we consider a lattice with periodic boundary conditions to efficiently extract and focus on the main new properties emerging from the model. Then we analyse the more realistic case of open boundary conditions, emphasise the equivalence of the emergent phenomenology with the periodic boundary case and we link our results to biologically relevant cases and potential experimental applications.

\subsection*{The two-state model}
To overcome the shortages of Poissonian-particle transport models, here we describe motor enzymes as interacting particles endowed with an internal kinetic cycle moving on a lattice.
For the sake of simplicity, we deal with the simplest enzymatic cycle illustrated in [\ref{eq::michaelis}], and we consider the backward rate $k_-$ negligible with respect to the forward rate $k_+ [S]$;  then we rename $k_+$ by $k$. This is a good approximation for many biological systems, such as coarse grained representation of the ribosomal biochemical cycle~\cite{wen_following_2008, ciandrini_ribosome_2013}.
We then investigate the collective movement of particles with this minimal stepping cycle: an irreversible substrate-dependent kinetic step and a translocation step defining a {\it two-state model} (see Fig.~\ref{fig::model})~\cite{ciandrini_role_2010}. A particle on the site $i$ of the discrete lattice makes a transition to an active, or {\it excited}, state with rate $k[S]$, and then translocates with rate $\gamma$ to the following site only if this is unoccupied. After the translocation the particle returns to its initial, {\it ground} state. As a result of the internal kinetics, particles can progress in their stepping cycle also when spatially blocked by their subsequent neighbours. 
The average residence time of a single isolated particle is then given by the sum of the average time needed to make the transition to the active state and the average time needed for translocation
$\langle \tau \rangle = 1/k[S]+1/\gamma=([S]+K_M)/[S]\gamma$.
Hence, we recover the Michaelis-Menten equation for the inverse of the average residence time, where
$\gamma$ represents the maximal velocity of an isolated motor and the parameter $K_M= \gamma/k$ is the Michaelis constant. With these prescriptions the hopping rate $p$ of a Poissonian particle defined in the previous section and the average hopping rate of an isolated two-state particle coincide: $p=1/\langle\tau\rangle$. This choice will then allow us to directly compare the two-state model to a standard exclusion process with hopping rate $p$.

A relevant parameter in this analysis is the ratio $w\equiv k[S]/\gamma$ that corresponds to the substrate concentrations in units of the Michaelis-Menten constant $K_M$. Changing $w$ then is equivalent to account for different substrate availabilities, from limiting ($w \ll 1$) to saturating ($w \gg 1$) conditions.

\subsection*{Internal molecular kinetic steps enhance collective transport} 
We now numerically simulate the rescaled current $J/p$ versus the density $\rho$ of particles for the two-state model and show how the parameter $w$ influences the current. We refer the reader to Appendix A for a detailed description of the simulation scheme.

We show that in the two-state model the rescaled current is strongly affected by the parameter $w$ controlling the substrate concentration (see Fig.~\ref{fig::comparison_current}A, blue and yellow circles).
This is in stark contrast to the collective dynamics of Poissonian particles, which are characterised by a
rescaled current $J/p$ independent of the substrate concentration (dashed line). Remarkably, the transport is substantially enhanced by the presence of the internal degree of freedom representing the stepping cycle (Fig.~\ref{fig::comparison_current}A). This behaviour is displayed for a broad range of the parameter $w$ and, as expected, when $w\rightarrow\infty$ the two models coincide.
Furthermore, Fig.~\ref{fig::comparison_current}A shows that, when the substrate concentration is severely limiting ($w\ll1$), $J(\rho)$ is no longer symmetric and, interestingly, the transport flux is optimised for high densities. Additionally, the velocity $v(\rho)$ is no longer linear with the density as in the case of Poissonian particles, and its dependence on $\rho$ is strongly affected by the substrate availability (Fig.~\ref{fig::comparison_current}B). 

Mean-field theories have attempted to analytically describe collective models of particles with this minimal stepping cycle~\cite{ciandrini_role_2010, klumpp_effects_2008}. However, they exhibit serious disagreement with numerical simulations, particularly in the regime $w\ll1$. Although in this paper we do not intend to provide a refined mean-field theory, we show that these deviations arise from emergent correlation effects that also influence substantially the randomness of tracer particles.

\subsection*{Residence time fluctuations of multiple particles with internal kinetic steps} 
\begin{figure}[bh]
	\begin{center}
	\includegraphics[width=0.45\textwidth]{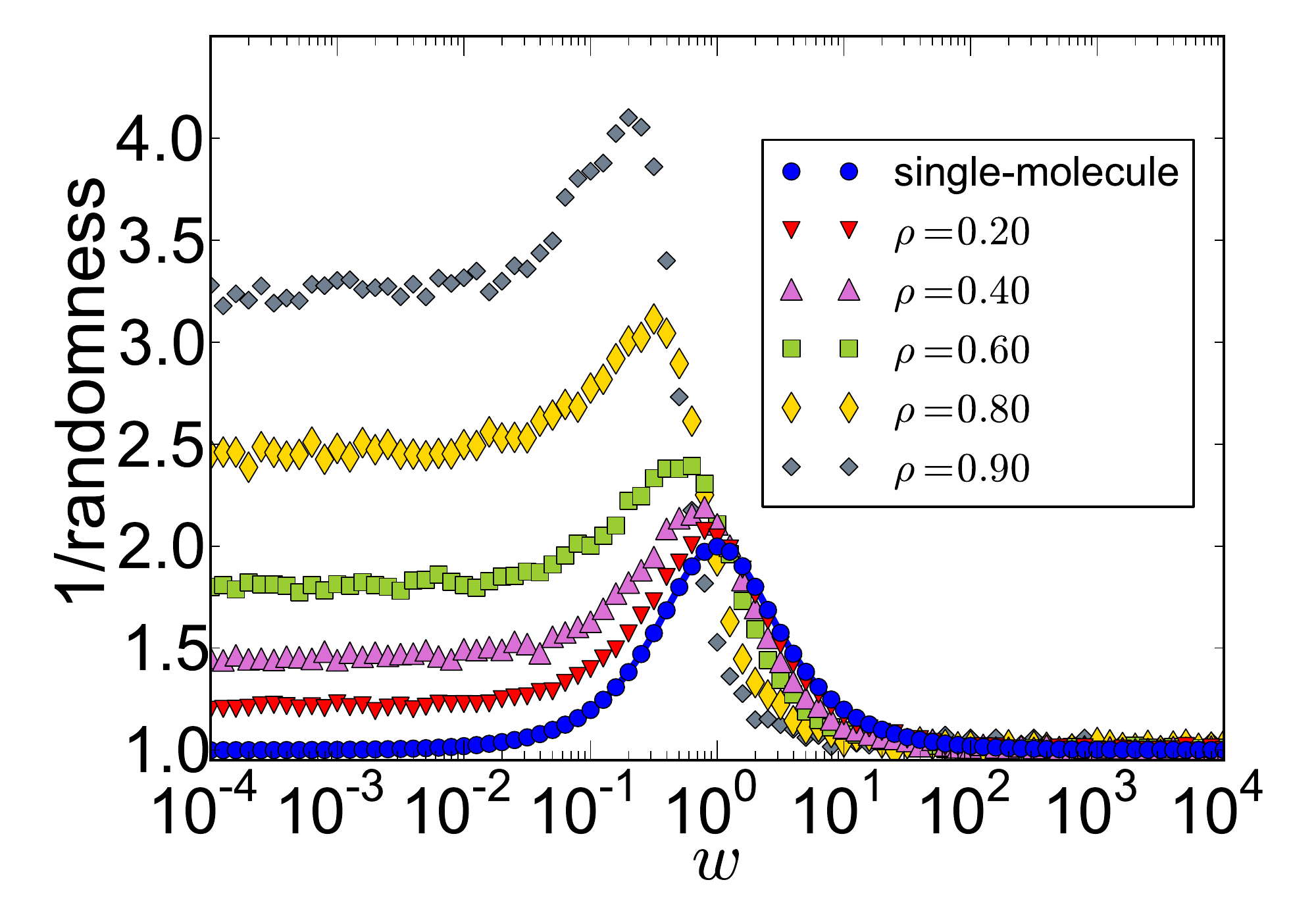}
	\caption{Inverse of the randomness $n=1/r$. Different colours and symbols represents systems at different densities (from $\rho=0.20$ to $0.90$). The blue circles instead represents a system composed of a single molecule, and the blue line (underneath the blue circles) is the analytical value of $n= (1+w)^2/(1+w^2)$ obtained from Eq.~[\ref{eq::randomness}]. The value of $n$ for an effective TASEP with hopping rate $p$ would instead be a constant ($n=1$), for any density. 
\label{fig::n}}
	\end{center}
\end{figure}

After having discussed average steady-state quantities like the current and the velocity, we now turn our attention to the fluctuations of the residence time of a tracer particle. In particular, we focus on the inverse of the randomness parameter, denoted by $n=1/r$. Notice that this quantity is no longer able to identify mechanistic constraints on the stepping cycle as in the single-molecule case~\cite{moffitt_methods_2010,moffitt_mechanistic_2010, chowdhury_stochastic_2013}. Figure~\ref{fig::n} shows the quantity $n$ measured for a tracer particle on a lattice with different overall particle densities $\rho$, for the two-state kinetics presented above. From these results it becomes evident that the randomness, and hence $n$, depends on the crowding of the lattice. First, the maximum of $n$ is no longer bound by the total number of kinetic steps of the motor (which would be equal to 2 in this case) as it occurs in the single-particle case~\cite{moffitt_methods_2010,moffitt_mechanistic_2010, chowdhury_stochastic_2013} (Fig.~\ref{fig::n}, blue circles). Second, the values of $n$ for limiting substrate concentrations, i.e. for $w\rightarrow 0$, do not give an indication of the substrate-binding steps, and they remarkably depend on the density $\rho$. In the saturating substrate concentration, for $w\rightarrow\infty$, which corresponds to the single state TASEP-like limit (the internal state has a very short lifetime, thus it is negligible), $n$ tends towards 1, as in a system of Poissonian steppers. 
As a result, the meaning of $n$, well defined in the single-molecule case as the number of rate limiting steps of the mechanochemical cycle, needs to be revised when considering collective transport of motors: it does not capture only features of the single-molecule kinetics, but it can provide information on the crowding and clustering of the system, as we show in the following sections. Accordingly, our analysis also shows that relevant information on the local density of motors around the tracer can be extracted by measuring its randomness $r$.  

We stress once more that when considering Poissonian particles, one obtains $n=1$ independently from the substrate abundance, since varying $p$ only rescales the timescale and does not affect the stochasticity of the system. Instead, by considering an additional kinetic state, the particle residence time is no longer exponentially distributed (as in a stochastic Poissonian process), but the distribution can be interpreted in terms of a convolution of exponentials with typical times related to the microscopic rates of the stepping cycle of the motor~\cite{kou_single-molecule_2005, sharma_distribution_2011}. 
If only part of the kinetic cycle is influenced by the presence of other enzymes on the lattice (as in the two-state model, where only the translocation step is affected by the density of enzymes), the residence time distribution can be strongly conditioned by the neighbourhood occupancy, giving rise to long range spatial correlations that we study in the next section.

\begin{figure*}[]
\begin{center}
\includegraphics[width=1\textwidth]{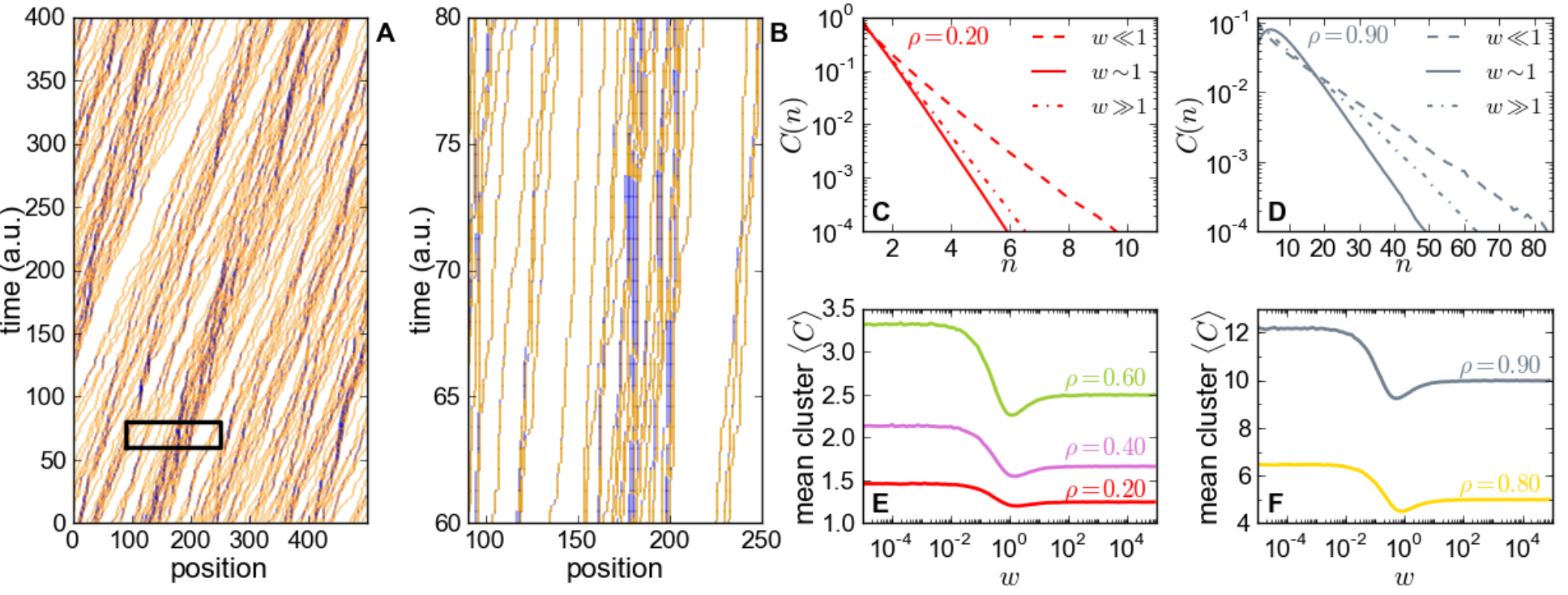}
\caption{Clustering of processive enzymes is tuned by the substrate availability. Panel {\bf A} shows the kymograph of a system with substrate-limiting conditions ($w=0.01$) and the region delimited by the black lines is magnified in panel {\bf B}. Active particles are coloured in blue while ground state particles are shown in yellow. It is evident the formation of cluster of particles and shocks caused by the internal kinetics. Panels {\bf C} and {\bf D} show the distributions of clusters $C(n)$ for $\rho=0.20$ and $\rho=0.90$, respectively, in three different cases: $w\ll1$, $w\sim1$, $w\gg1$. When the substrate is limiting ($w\ll1$) the cluster distributions display large tails, while when $w\sim1$ the distributions are less disperse in favour of smaller clusters. The non-monotonous behaviour of the mean of the cluster distributions is shown in panels {\bf E} and {\bf F} for different densities. The minimum of $\langle C \rangle$ is reached at the value of $w$ corresponding to the maximum of $n$ in Figure~\ref{fig::n}.\label{fig::big}}
\end{center}
\end{figure*}
\subsection*{Clustering induced by kinetic steps and detecting technique} 
A closer look at the system via kymographs (snapshots of the system at consecutive times tracing the trajectories of the particles, Fig.~\ref{fig::big}A-B) reveals the emergence of clusters and queues of active particles for substrate-limiting conditions ($w=0.01$) and particle density $\rho=0.2$. A more quantitative analysis shows that by changing $w$, and therefore the substrate dependence, clustering increases when the substrate is limiting and the probability of finding large clusters is no longer negligible. This is clearly shown by computing the cluster distribution $C(n)$, i.e. the probability of finding a cluster of particles of size $n$ ($n$ adjacent particles). In Fig.~\ref{fig::big} we show $C(n)$ versus $n$ for different values of $w$ for $\rho=0.20$ (C) and $\rho=0.90$ (D): this probability distribution displays large exponential tails in substrate limiting conditions, i.e. for small values of $w$. It also shows a minimal dispersion when $w$ is of order 1, and then the large-cluster tail moderately increases for larger values of $w$ (i.e. in the single state TASEP-like limit).

As a consequence of the peculiar features of $C(n)$, the mean cluster size $\langle C \rangle$ arising from this distribution exhibits a non-monotonous dependency on the substrate concentration $w$, as illustrated in Figure~\ref{fig::big}E-F.  Since the broadness of $C(n)$ becomes smaller around $w=1$ (see Fig. \ref{fig::big}C), the mean cluster size as a function of $w$ exhibits a minimum when $w\simeq 1$, i.e. the average cluster is smaller than in both the substrate limiting and saturating (TASEP-like) cases.
 
\begin{figure}[h]
\begin{center}
\includegraphics[width=0.5\textwidth]{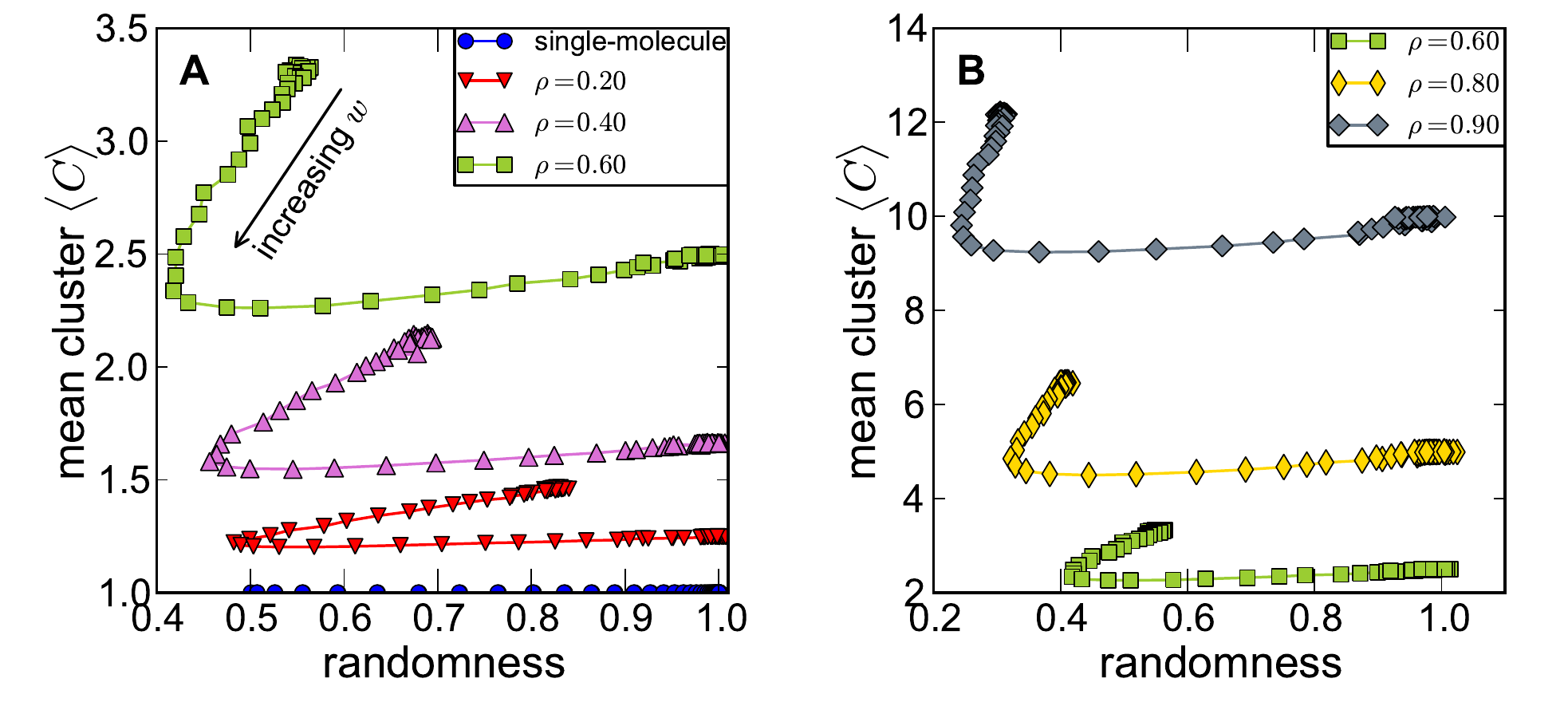}
\caption{Mean cluster as a function of the randomness for the single-molecule case (blue circles) and densities $\rho=0.20, 0.40, 0.60$ ({\bf A}) and $\rho=0.60, 0.80, 0.90$ ({\bf B}). By decreasing the relevance of the substrate-depending internal states (i.e. by increasing $w$) one advances on the cusp-shape curves starting from large mean cluster values as indicated by the black arrow in panel {\bf A}. Despite of the non-univocal nature of this relationship, for small values of the randomness the cluster-size can be indirectly measured by observation of the randomness and knowledge of the substrate conditions. \label{fig::randomness_cluster}}
\end{center}
\end{figure}
In the previous section we have shown that the inclusion of the enzymatic kinetic cycle, even in its simplest form, induces localized inhomogeneities of the density such as particle crowding and queuing. Due to the strong non-monotonous dependence of both the randomness parameter (Fig.~\ref{fig::n}) and the mean cluster size (Fig.~\ref{fig::big}E-F) on $w$, and therefore on the substrate abundance, it is interesting to plot the mean cluster size $\langle C\rangle$ versus the randomness parameter (Fig.~\ref{fig::randomness_cluster}). 
The non-monotonous dependence of the randomness on $w$ implies that by starting from very small values of $w$ and increasing it, one would eventually meet a minimum of the randomness (maximum of $n$, see Fig.~\ref{fig::n}), then see the fluctuations increasing again and saturate to 1 with the type of kinetics analysed here. The non-trivial mean cluster size dependence on $w$ shows a similar behaviour, and by merging this information for each density one obtains the cusp-shaped curves presented in Fig.~\ref{fig::randomness_cluster}.  Hence, experimental measurements of the randomness could help identifying different clustering regimes at different substrate concentrations, as Fig.~\ref{fig::randomness_cluster} shows. 
Strikingly, the minimum of the randomness corresponds to the minimum of the mean cluster size, tracing a link between the fluctuations of a tracer particle and the collective properties of the system. The cluster formation mechanism at low substrate seems to decrease the stochasticity; in other words, the randomness in the limit $w \ll 1$ is smaller than the single-molecule case (see Fig.~\ref{fig::n}).

\subsection*{Open Boundary conditions}
In this section we study the consequences of considering the model presented above with open boundary conditions, i.e. without fixing the density of particles but allowing particles to enter from one end of the lattice with rate $\alpha$, and exit from the other end with rate $\beta$. As it is well known from the TASEP-literature, this model presents boundary induced phase transitions~\cite{krug_boundary_1991}. When the entry rate $\alpha$ is limiting the system is found in a low density (LD) phase, while when the exit rate $\beta$ constitutes a bottleneck the lattice is in a high density (HD) phase characterised by a queue of particles piling up from the end of the lattice. Since exclusion interactions limit the current of particles in the bulk, a maximal current (MC) phase is then observed. This is characterised by a smaller density compared to the HD phase and, as the name suggests, the largest current of particles  flows through the lattice (corresponding to the maximum of the curve $J(\rho)$, see Fig.~\ref{fig::comparison_current}). By then changing the entry and exit rates it is possible to move between these different phases. For the features of the different phases and analysis of the rich phase diagram of TASEP-like models we refer the reader to other publications~\cite{chou_non-equilibrium_2011}. For our purposes it is useful to notice that, although the density of particles $\rho$ is no longer a system control parameter, the average occupation $\bar\rho$ of the lattice is determined by fixing the rates $\alpha$ and $\beta$, allowing a mapping between the close and the open boundary cases. The same argument holds for the two-state model~\cite{ciandrini_role_2010}. The average density of particles on the lattice is regulated by the triplet $\alpha/\gamma, \beta/\gamma, w$, see also the phase diagram of the system (Fig. S1 in the Supporting Material). Importantly, for a fixed pair $\alpha/\gamma, \beta/\gamma$, by decreasing $w$ the open system eventually undergoes a phase transition towards the MC (Figure S1). 

Hence, if in the periodic case we isolated the effects that the stepping cycle has on the dynamics of the motors and on the local clustering for a fixed density, by changing $w$ in the open case we also to take in account the differences in the density induced by the boundary-controlled phase transitions. Despite that, the outcomes can still be interpreted within the same framework, and one is still able to provide estimates of the crowding of the filament from measurements of the randomness parameter. We show that with some representative examples. 

\begin{figure}[h]
\begin{center}
\includegraphics[width=0.5\textwidth]{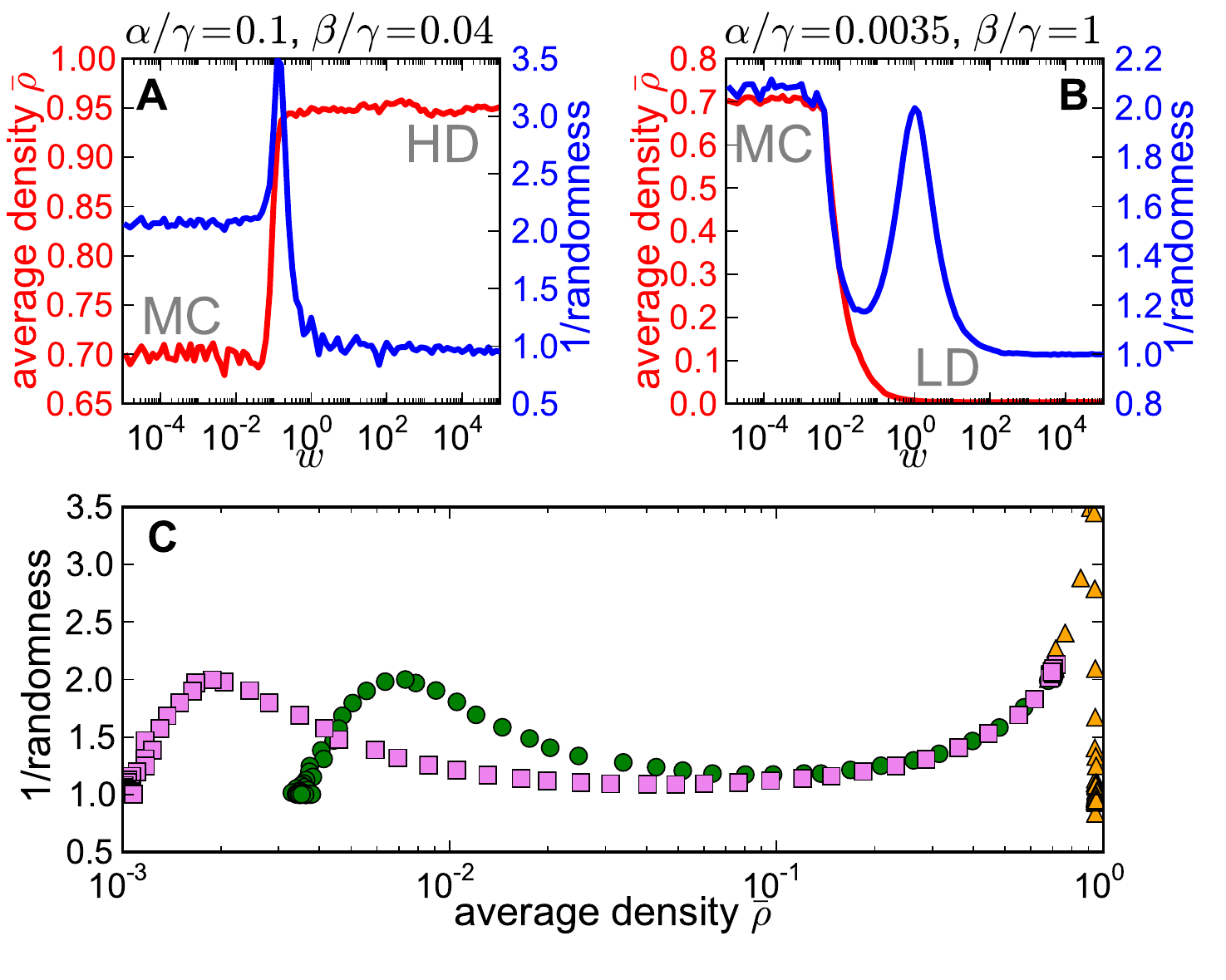}
\caption{Simulations in the open boundary case. Panels ({\bf A}) and ({\bf B}) show the average density $\bar\rho$ (in red) and inverse of the randomness (in blue) as a function of the parameter $w$ for different pairs of entry and exit rates. Panel ({\bf C}) represents the randomness parameter as a function of the average density $\bar\rho$. Green circles correspond the simulations presented in ({\bf B}), i.e. $\alpha/\gamma = 0.0035$ and $\beta/\gamma=1.0$, pink squares are for $\alpha/\gamma = 0.001$ and $\beta/\gamma=0.6$ and orange triangles with $\alpha/\gamma = 0.1$ and $\beta/\gamma=0.04$. \label{fig::open}}
\end{center}
\end{figure}

As a first example we consider a crowded situation at high substrate concentrations, i.e. with the system found in the HD when $w$ is large, as shown in Figure~\ref{fig::open}A. By decreasing $w$ the inverse of the randomness $n$ remains constant at the beginning, then the dependence shows a steep increase when $w\sim1$, as shown also in Figure~\ref{fig::n} for comparable densities (gray diamonds) in periodic boundary conditions. The system then undergoes a transition to the MC phase\footnote{Note that, contrary to the TASEP, in the two-state model the density in the MC phase is in general different from 0.5, see Ref.~\cite{ciandrini_role_2010}}, causing a change in the average density $\bar\rho$. Once the system is deep in the MC phase its average density is smaller than in the HD regime, and the randomness computed corresponds to the one calculated with a closed system for similar density at limiting substrate concentrations ($\bar\rho\sim0.7$, between the green squares and yellow diamonds in Figure~\ref{fig::n}).
The peak in $n$ that is present in the open boundary case is induced by the stepping cycle and it has hence the same nature of the one observed in the close boundary case. We in fact observe a reduction of the average particle cluster size in correspondence to the peak in $n$ and before the transition to the MC (see Fig. S2 in the Supporting Material). We note that this fact provides evidence that the periodic lattice case previously studied can already capture many of the features of the open boundary case. 

The second example consists of a lattice in the LD phase when the substrate is present at high concentrations, Figure~\ref{fig::open}B. As a result of the small entry rate (compared to the other parameters), the average density $\bar\rho$ of motors on the track is small and the exclusion interactions can at a first approximation be neglected for large $w$. The plot of $n$ then resembles the one of the single molecule in the close boundary condition (blue circles in Figure~\ref{fig::n}), and it presents a maximum at values $w\sim1$. However, when $w$ decreases the lattice enters the MC regime, the density effects become evident and the inverse of the randomness increases again (generating a minimum in $n$) to reach the values expected by the periodic boundary simulations. 

These two examples represent two biologically relevant situations: the first case is representative of processive motors that are slowed down by a bottleneck situated at the exit, and the second one of motors that are in excess of substrate (e.g. high ATP concentrations) (large $w$) in the LD phase. In particular, the second case can be applied to ribosomes advancing on an mRNA. Indeed, in order to produce Figure~\ref{fig::open}, we have used biologically realistic parameters for the rates $\alpha$ and $\beta$~\cite{ciandrini_ribosome_2013}. The parameter $w$ in case of mRNA translation is mainly given by the abundance of cognate tRNA, and is smaller than 1 ($10^{-2} \lesssim w \lesssim 1$ in \textit{S.cerevisiae})~\cite{ciandrini_ribosome_2013}. From Figure~\ref{fig::open}B one can then speculate that $w$ might have an important role in regulating the randomness of the ribosomes traffic and therefore the noise in protein synthesis at the level of translation. We are aware that inhomogeneities in the sequence could further influence the randomness of ribosomes. Here we consider the case in which the nucleotide sequence does not have relevant bottlenecks, so that the dynamics can be considered, in first approximation, uniform along the lattice. In this work we have decided to only treat the homogeneous case to better separate the effects of the lattice inhomogeneities and the (more counterintuitive) ones induced by the motor's stepping cycle. 

As in the periodic boundary case one could imagine to extract information on the filament crowding by the measurement of the randomness of a tracer. In Figure~\ref{fig::open}C we have therefore represented the randomness as a function of the average density $\bar\rho$. Interestingly, other than direct measurements of the randomness of a molecular motor by, e.g. quantum dots~\cite{seitz_processive_2006, courty_tracking_2006}, in the case of protein synthesis one can also relate this quantity to the noise in gene expression.

The results presented above are valid for highly processive motors entering on one side of the track and being depleted at the other end. However, in order to consider finite processivity one should consider attachment and detachment of motors at any position along the lattice~\cite{parmeggiani_phase_2003}. While we would expect the main phenomenology presented here to hold with those extensions (the randomness depending on local density), in order to provide quantitative predictions one should adapt the properties (number of states and transition rates) of the stepping cycle to the case of interest. Moreover, if considering finite processivity motors, the detachment could depend on the biochemical state of the motor~\cite{crevel_what_2004, telley_obstacles_2009}; we expect that this can give rise to many variations of the model whose analysis is out of the scope of our present work.

\section*{DISCUSSION}
The several biochemical reactions involved in a single move of an enzyme define its kinetic cycle, which has been thoroughly studied at a single-molecule level both theoretically~\cite{kolomeisky_molecular_2007, moffitt_methods_2010,moffitt_mechanistic_2010} and experimentally~\cite{schnitzer_kinesin_1997,visscher_single_1999}. Despite significant progress of single-molecule techniques, how the internal stepping dynamics affects collective transport of molecular motors still remains an outstanding open question in molecular and cellular biology.

In this work we have unveiled features of collective enzyme kinetics and shown emergent (temporal and spatial) correlations induced by internal molecular kinetics, with a particular emphasis on biologically driven transport process by processive enzymes.

We have thus focused on the collective properties of processive enzymes on the same biological track (e.g. motor proteins on a cytoskeletal filament) and first shown that collective models of unidimensional transport such as the Totally Asymmetric Simple Exclusion Process (TASEP) do not capture the complexity and the interplay between the individual particle's stepping cycle and the macroscopic phenomenology of the system. Models built on single particle steps without internal states (unstructured Poissonian walkers) are characterised by stationary states that neither catch the well known dependence on the substrate concentration of, for example, the fluctuations of the particles' residence times (the randomness), nor exhibit the rich phenomenology associated with the fluctuations of the residence time and also the emerging spatial correlations.

To bridge the gap between single-molecule and multi-particle models we have extended a prototypical unidimensional transport process (the TASEP) by explicitly incorporating an intermediate state defining a stepping cycle on each particle.
The inclusion of the enzymatic stepping cycles to a simple driven lattice gas model has now allowed us to study the role of the kinetic cycle (and, e.g., its substrate dependence) at a collective scale (see, e.g.,~\cite{klumpp_effects_2008, garai_fluctuations_2009, ciandrini_role_2010}). As a proof of principle we have studied  the simplest kinetic cycle, a Michaelis-Menten-type with an irreversible substrate dependent transition and a translocation step (which can be easily generalised to more complex kinetics schemes). This unsophisticated stepping cycle is sufficient to give rise to complex dynamical phenomena that reproduce the phenomenology of experimental results. Moreover, our approach is also justified by the presence of two major timescales in the stepping in many molecular motors~\cite{rief_myosin-v_2000, yildiz_kinesin_2004,wen_following_2008}.

We have first highlighted and revisited previous results: (i) the stepping cycle can enhance the current of particles if compared to the standard TASEP, (ii) the optimal transport capacity is obtained for higher densities, and (iii) at low substrate abundance the velocity $v(\rho)$ of a tracer particle depends in a nonlinear way on the density.
These predictions could be verified by {\it in vitro} and {\it in vivo} experiments of processive molecular motors at low ATP concentrations, similar to the ones performed in~\cite{seitz_processive_2006, courty_tracking_2006, leduc_molecular_2012}.

Furthermore, the phenomenology observed in the experiments, including the substrate-dependent fluctuations of the residence times~\cite{schnitzer_kinesin_1997,visscher_single_1999}, is recovered by the inclusion of stepping cycle kinetics. Here we have shown that the randomness of a tracer particle strongly depends on the crowding of the system, and significantly deviates from the single-molecule behaviour. We immediately noticed that particles are no longer characterised by Poissonian exponential residence times, whose distribution can be expressed by a convolution of exponentials. These distributions also bear some similarities to the gamma waiting times distributions characteristic of semi-Markov processes; the fact that only part of the stepping cycle is influenced by the crowding of the system can indeed be seen as a peculiar memory property of the individual particles~\cite{gorissen_ribosome_2012}.

Thus, in contrast to the case of an isolated molecule, the inverse $n$ of a randomness parameter does not give the number of limiting steps of the enzyme kinetics anymore. Interestingly, we have found that the crowding remarkably affects measurements of the randomness, in particular at low and medium substrate concentrations, and $n$ is no longer bound by the number of kinetic steps.

In the periodic system we have focused on the time evolution of the system via the kymographs and obtained signatures of relevant particle clustering at low substrate abundance. We have shown that when the system lacks the resources necessary to force a particle to the excited state ($w \ll 1$), the limiting substrate-dependent rate has the remarkable effect of generating moving and spatially localised queues of particles (clusters). This is consistent with the results previously discussed: particles trapped in a queue can in fact use the time to undergo a transition to the active state, and then move almost simultaneously, producing a larger current compared to standard exclusion processes. Clustering at low $w$ also explains why $n$, effectively interpreted as the number of limiting steps, is larger than 1 in such conditions. In fact, a cluster of particles can be thought of as an extended object that necessitates more than one transition before undergoing a translocation step.
Moreover, from a theoretical point of view, we stress that the mean cluster size $\langle C \rangle$ remains finite, and there is no transition towards a unique extensive cluster as displayed by other models~\cite{oloan_jamming_1998, nishinari_cluster_2003}. 

In order to make a more appropriate connection to biological systems, we have also analysed the model in the open boundary case. Although the model in open boundaries also presents phase transitions influencing the randomness and the crowding of the system, the framework and the results obtained in the periodic case remain valid. Since a given set of entry and exit rates determines the average density on the track, we can roughly map the open system to the periodic one. Further extensions are out of the scope of the present work.

On biological grounds, we present a clear mechanism that allows processive enzyme like motor proteins to spontaneously form clusters at low concentrations of energy resources like ATP, and at the same time to improve the overall transport at high densities. In addition, we have shown that there is a clear link between randomness measurements and local density, and therefore suggest that the evaluation of the randomness by tracer particles might be used as a prompt to detect the crowding on the track. Using an individual tracer might constitute an alternative strategy to provide quantitative density measurements. This methodology is particularly relevant in crowded regimes in which the relation between fluorescence and motor density is not linear. 
The same arguments can be used to many biological tight-coupled processive machines, as ribosomes translating an mRNA, which are known to usually work in the regime $w \ll 1$ (here the substrate is mainly determined by the abundance of cognate tRNAs). In this case the randomness contributes to noise in gene expression, which might be used as an indirect measure.
This study opens to the interpretation of tracing particle statistics of experiments {\it in vivo} and {\it in vitro}~\cite{seitz_processive_2006, courty_tracking_2006} and it allows us to bridge the gap between refined single-molecule theories (e.g. ratchet models~\cite{kolomeisky_molecular_2007}) and coarse-grained cooperative models of molecular motors, which are recovered in particular conditions, in order to understand the principles of motor collective coordination. 

\begin{acknowledgments}
The authors would like to thank I.~Stansfield for the useful discussions. LC is supported by an EMBO long-lerm fellowship co-funded by the European Commission (EMBOCOFUND2010, GA-2010-267146). AP is financially supported by the Scientific Council of the University of Montpellier 2 and the Laboratory of Excellence NUMEV. MCR thanks Biotechnology and Biological Sciences Research Council (BB/F00513/X1) and SULSA for financial support. LC would also like to thank the ICSMB at the University of Aberdeen for the computational time support.
\end{acknowledgments}

\section*{APPENDIX A: numerical simulations}
We have performed numerical simulations by implementing a continuous time Monte Carlo algorithm based on the Gillespie algorithm~\cite{gillespie_exact_1977}. Individual particles on a discrete lattice undergo a transition to their active state with a rate $k[S]$, then move with rate $\gamma$ provided that the next site is not occupied by another particle. We start collecting data once the system has entered the steady-state. We typically neglected the first $2\cdot10^6$ runs of our algorithm, and then collected data for other $2\cdot10^6$ runs. Without loss of generality, we fix $\gamma=1$ s$^{-1}$ and changed the transition rate to achieve different values of $w$.
A cluster of size $n$ is defined as a group of $n$ adjacent particles. To compute the cluster-size distributions and the density we weighed the particle configurations with the Gillespie intervals $dt$. We compute the dwell time distribution of tracer particles and compute the randomness by using Eq.~[\ref{eq::randomness}].

For the periodic case, the figures in this paper (Figure 2,3,4 and 5) show the outcome of simulations for a periodic lattice composed of 500 sites, and densities as displayed in the legends. In the open boundary case (Figure 6), to better avoid finite size effects we instead considered a lattice of length $L=1000$ sites.



\ 
\newpage

\setcounter{figure}{0}
\makeatletter 
\renewcommand{\thefigure}{S\@arabic\c@figure}
\makeatother

\begin{figure*}[h!]
\begin{center}
\section*{SUPPLEMENTARY FIGURES}
\includegraphics[width=0.4\textwidth]{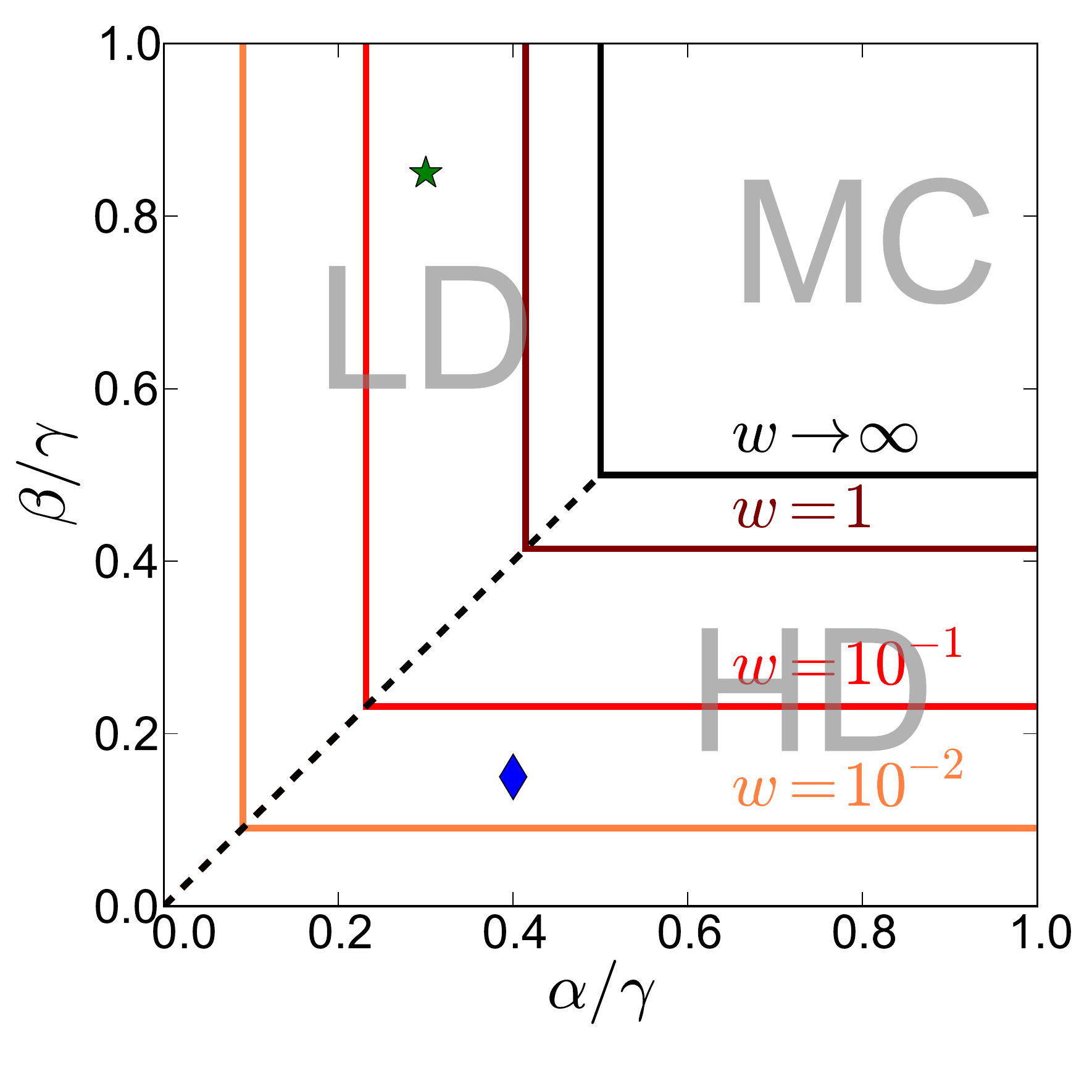}
\caption{Phase diagram of the two-state model as presented in~\cite{ciandrini_role_2010}. The three phases (LD, HD, MC) can be reached by varying the boundary rates $\alpha$ and $\beta$, the translocation rate $\gamma$ or the parameter $w$. By decreasing $w$ the MC region takes a larger and larger part of the phase diagram, while in the limit $w\rightarrow \infty$ the well known results from TASEP are recovered, with transition lines at $\alpha/\gamma=\beta/\gamma=0.5$. Starting from a point in the phase diagram in the HD or LD (green star or blue circle) as done in Fig.~\ref{fig::open}A and B, by decreasing $w$ the MC phase is eventually reached. \label{fig::S1}}
\end{center}
\end{figure*}
\ 
\begin{figure*}[h!]
\begin{center}
\includegraphics[width=0.4\textwidth]{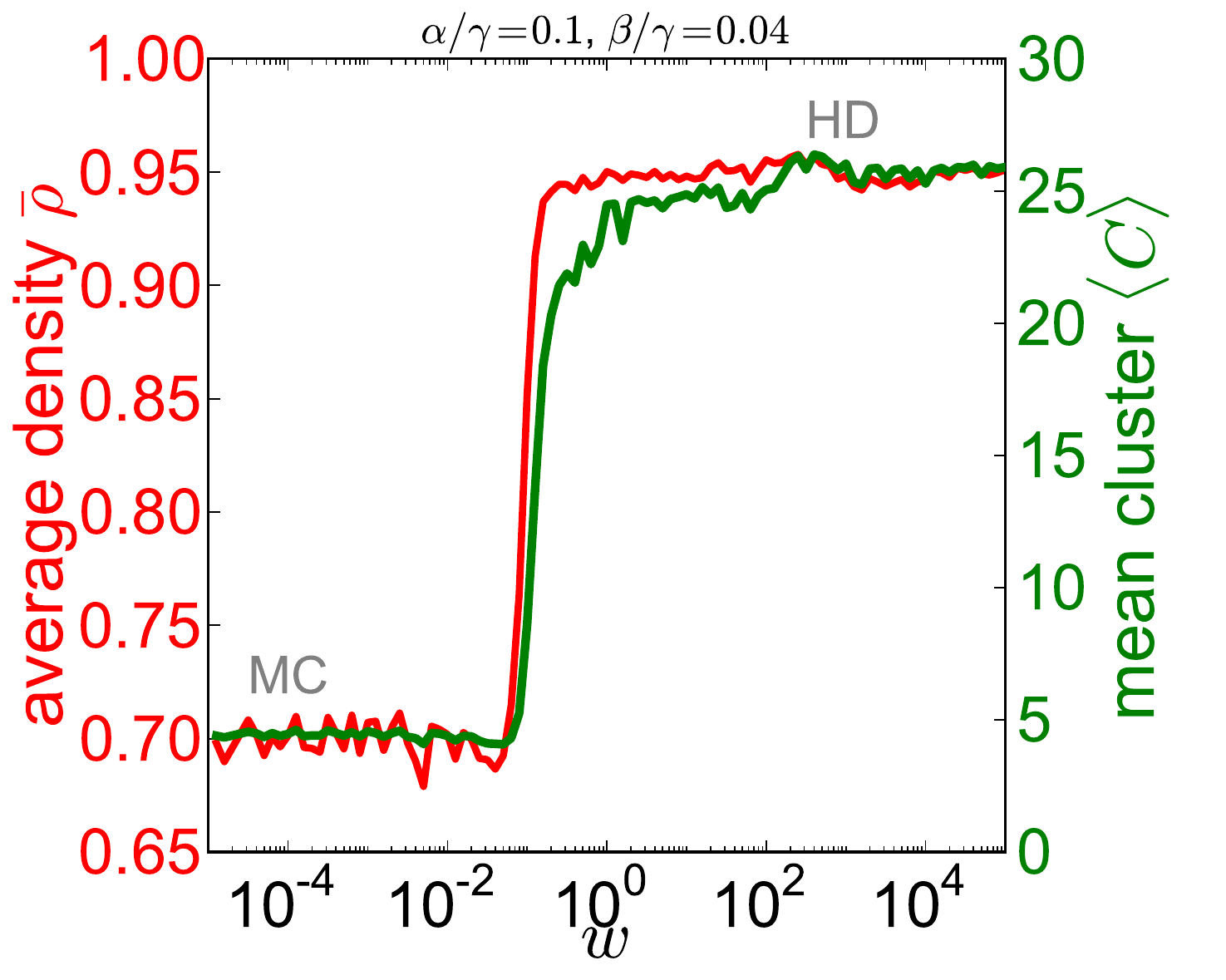}
\caption{Average density $\bar\rho$ and mean cluster size $\langle C \rangle$ as a function of $w$ with $\alpha/\gamma =0.1, \beta/\gamma =0.04$ as shown in Fig.~\ref{fig::open}A. By decreasing $w$ we observe a reduction of the mean particle cluster size (green line) before the drop in $\bar\rho$ indicating the transition to the MC (red line). The reduction in $\langle C \rangle$ is an effect of the kinetic step, as explained in the periodic boundary case, and it explains the peak in $n$ (Fig.~\ref{fig::open}A).
In fact, before the transition the average density is approximatively constant and the system can be rationalised with the known results of the periodic boundary. \label{fig::S2}}
\end{center}
\end{figure*}

\end{document}